\documentclass[journal]{IEEEtran}

\usepackage{cite}
\usepackage[pdftex]{graphicx}
\usepackage{amsmath, bm, amssymb}
\usepackage{algorithmic}
\usepackage{array}
\usepackage{subcaption}
\usepackage{xcolor}

\def\j{\mathrm{j}}
\def\e{\mathrm{e}}
\def\c{\mathrm{c}}
\def\x{\mathrm{x}}
\def\y{\mathrm{y}}
\def\p{\mathrm{p}}
\def\f{\mathrm{f}}
\def\b{\mathrm{b}}
\def\eff{\mathrm{eff}}

\def\ve#1{{\mathchoice{\mbox{\boldmath$\displaystyle #1$}}%
{\mbox{\boldmath$\textstyle #1$}}%
{\mbox{\boldmath$\scriptstyle #1$}}%
{\mbox{\boldmath$\scriptscriptstyle #1$}}}}

\begin{document}

\title{Joint PMD Tracking and Nonlinearity Compensation with Deep Neural Networks}

\author{Prasham~Jain,
        Lutz~Lampe,
        and~Jeebak~Mitra
        \thanks{Manuscript updated on \today.}
\thanks{P. Jain and L. Lampe are with the Department
of Electrical and Computer Engineering, University of British Columbia, Vancouver,
BC V6T 1Z4, Canada, e-mail: (prashamj@ece.ubc.ca, lampe@ece.ubc.ca).}
\thanks{J. Mitra was with Huawei Technologies Canada, Ottawa, He is now with Dell Technologies Canada, e-mail:  jeebak.mitra@dell.com}}

\markboth{Jain \MakeLowercase{\textit{et al.}}:Joint PMD Tracking and NLC with Deep Neural Networks}{}

\maketitle

\begin{abstract}
Overcoming fiber nonlinearity is one of the core challenges limiting the capacity of optical fiber communication systems. Machine learning based solutions such as learned digital backpropagation (LDBP) and the recently proposed deep convolutional recurrent neural network (DCRNN) have been shown to be effective for fiber nonlinearity compensation (NLC). Incorporating distributed compensation of polarization mode dispersion (PMD) within the learned models can improve their performance even further but at the same time, it also couples the compensation of nonlinearity and PMD. Consequently, it is important to consider the time variation of PMD for such a joint compensation scheme. In this paper, we investigate the impact of PMD drift on the DCRNN model with distributed compensation of PMD. We propose a transfer learning based selective training scheme to adapt the learned neural network model to changes in PMD. We demonstrate that fine-tuning only a small subset of weights as per the proposed method is sufficient for adapting the model to PMD drift. Using decision directed feedback for online learning, we track continuous PMD drift resulting from a time-varying rotation of the state of polarization (SOP). We show that transferring knowledge from a pre-trained base model using the proposed scheme significantly reduces the re-training efforts for different PMD realizations. Applying the hinge model for SOP rotation, our simulation results show that the learned models maintain their performance gains while tracking the PMD.
\end{abstract}


\IEEEpeerreviewmaketitle

\section{Introduction}
\IEEEPARstart{C}{apacity} of optical fiber communication systems is limited by the non-linear Kerr effect \cite{kerr}. During signal propagation, this nonlinearity interacts with linear effects such as chromatic dispersion (CD) \cite{cd} and polarization mode dispersion (PMD) \cite{pmd}. Each impairment in isolation can be compensated with relatively simple solutions by solving the associated special case of the non-linear Schr\"odinger equation (NLSE). The optical receiver, typically composed of a cascade of digital signal processing (DSP) blocks, compensates each effect based on such model assumptions \cite{receiver}. However, it is challenging to develop a method which can effectively capture the interplay of nonlinearity with linear distortions and invert its effect. The split-step Fourier method (SSFM) \cite{ssfm} is a popular numerical method which solves the propagation equations by splitting the link into steps and applying the linear and non-linear distortions independently. The core  assumption of SSFM being that the distortions can be decoupled for a small enough step size. Among the conventional nonlinearity compensation (NLC) techniques, digital backpropagation (DBP) \cite{DBP} leverages the SSFM to solve the inverse propagation equations. However, DBP becomes prohibitively expensive for long-haul fiber due to the large number of steps required to maintain performance.

Recently, machine learning based methods have shown immense potential for NLC by learning the non-linear transfer function directly from data. It has been shown that learned NLC methods can be effective even without any a-priori knowledge of the fiber parameters \cite{CVNN}. Parameterization of existing models is an effective approach for developing learned NLC methods. The learned digital backpropagation (LDBP) method \cite{LDBP} has been developed by parameterizing the SSFM for the standard NLSE. For dual-polarized (DP) transmission, it has been extended to include distributed compensation of PMD \cite{ldbp-pmd} by parameterizing the SSFM for the Manakov-PMD equations. The resulting approach is referred to as LDBP-PMD. While LDBP presents a significant improvement over conventional DBP, it still shares limited capability to capture the interplay of dispersion and nonlinearity, given the large step size required to maintain feasibility. Recently, we proposed the deep convolutional recurrent neural network (DCRNN) model \cite{jain:2022} to overcome this limitation by using a bi-directional recurrent neural network to capture the interaction of nonlinearity and dispersion based on past (and future) symbols at each NLC step. Similar to \cite{ldbp-pmd}, its extension DCRNN-PMD incorporates the distributed compensation of PMD. 

However, inclusion of distributed PMD in the learned model results in the coupling of NLC with PMD compensation and thus, requires additional considerations since in practice, PMD may drift over time\cite{poole1997polarization}. Learned NLC schemes in the literature often assume that the model can be trained offline for an arbitrarily long period of time using a sufficiently large training dataset from a stationary channel/distribution \cite{DCNN}. But when we consider time evolution of PMD, the distribution of observed data becomes non-stationary, a phenomenon referred to as concept drift in the machine learning literature \cite{drift2}. For practical application, it is therefore essential that the learned model can adapt in real-time in the presence of concept drift using an online training scheme. 
Unfortunately, previous studies on adaptation of joint NLC and PMD compensation schemes have only considered instantaneous uncorrelated changes in the state of PMD, wherein re-training the converged model has been found to be equivalent to learning the PMD compensation parameters from scratch or worse \cite{ldbp-pmd}. 
Transfer learning has been shown to be an effective method to reduce the re-training overhead of learned NLC models \cite{FreireJLT:2021,Simeone:meta-learning}. Also, unsupervised learning in the form of K-means clustering has been applied to train the model online \cite{Giacoumidis:2020}. However, these studies have not investigated the impact of time-varying channel impairments. On the other hand, previous works on adapting a learned model to continuous changes in PMD have either ignored the presence of nonlinearity or assumed that it has been compensated by other means \cite{Schmalen:2022}, thereby, limiting the investigation to a purely linear channel.
In fact, real-time adaptation of learned NLC methods to time-varying channel impairments has not been investigated so far in the literature.

In this paper, we propose a transfer learning based selective online training scheme to efficiently transfer knowledge from an offline trained base model to the filter in operation. As per the proposed scheme, we only update the learned weights of the PMD compensation layers while leaving the remaining learned model unaffected. This approach builds upon the training method applied in \cite{ldbp-pmd}, where the learned coefficients from the LDBP model are frozen and serve as the initialization for the LDBP-PMD model. 
By limiting the trainable weights, the proposed approach greatly reduces the training overhead. We extend the method through incorporation of online learning, based on decision directed feedback, to perform lifelong real-time adaptation to changes in PMD.    
For concreteness, we apply the hinge model for time-evolution of the state of polarization (SOP) to simulate continuous drift in PMD \cite{hinge-model}. While previous attempts to adapt a pre-trained learned NLC model to a different realization of PMD found limited success \cite{ldbp-pmd}, 
 our results from system simulations demonstrate that knowledge can be effectively transferred and performance can be sustained, thus bridging the gap between extended offline training and real world application of learned NLC methods.

In addition to the new adaptive training scheme, combining the principles of transfer learning and online learning for continuous adaptation of the learned NLC model, this paper extends the conference version \cite{jain:2022} by providing a detailed construction of the DCRNN-PMD model, discussing various design choices and extending the numerical results to WDM transmission systems. 

The remainder of this paper is organized as follows. In Section~\ref{model-description}, we briefly review the theoretical background, including the hinge model for continuous SOP drift. The DCRNN-PMD model and the transfer learning based online model adaptation scheme for tracking PMD drift are developed in Section~\ref{s:DCRNN}. 
In Section~\ref{pruning}, we discuss the methodology applied to compare the performance-complexity trade-off of learned NLC models. 
In Section~\ref{results}, we provide details of our transmission system, the hyperparameters of the various compensation schemes and the NN training routine, followed by the performance and complexity results for various NLC techniques for both static and time-varying channels. Section~\ref{conclusion} provides concluding remarks for the paper.

\section{Modeling of Fiber Propagation Channel} \label{model-description}

In this section, we provide a brief review of the fiber model, which forms the basis for the design of the DCRNN-PMD model and provides support for the proposed adaptive training scheme. We also describe the hinge model, which is applied to simulate continuous PMD drift.

\subsection{SSFM Based Channel Model} \label{SSFM}
The propagation of a dual-polarized signal through a single-mode optical fiber, including birefringes and rotation of the principle states of polarization (PSP), can be described using the Manakov-PMD equation \cite{manakov-pmd} as
\begin{equation}
\begin{split}
       \frac{\partial \ve{E}(z,t)}{\partial z} = &- \frac{\alpha}{2} \ve{E}(z,t) + \Delta \beta_1 \ve{\Sigma}(z) \frac{\partial \ve{E}(z,t)}{\partial t}  \\&-\j \frac{\beta_2}{2} \frac{\partial^2 \ve{E}(z,t)}{\partial t^2} + \j \gamma \frac{8}{9} \ve{E}(z,t) \|\ve{E}(z,t)\|^2 ,
\end{split}
\end{equation}
where $\ve{E}(z,t) = [E_\x(z,t), E_\y(z,t)]^T$,  and $E_\x(z,t)$ and $E_\y(z,t)$ are the complex baseband signals of the X and Y polarizations, respectively. The signal is a function of propagation time $t$ and distance $0 \leq z \leq L$, where $L$ is the total length of the fiber. The fiber parameters include the attenuation coefficient $\alpha$, the group velocity dispersion coefficient $\beta_2$ responsible for CD, and the nonlinear coefficient $\gamma$. The coefficient $\Delta \beta_1 = (\beta_{1\x} - \beta_{1\y})/2$ represents the differential group delay (DGD) between the two polarizations along the PSP. In relation, the PMD parameter can be described as $2\sqrt{2L_\c} \Delta \beta_1$, where $L_\c$ is the correlation length of the two polarizations. The matrix $\ve{\Sigma}(z)$ represents the PSP rotation at distance $z$, which causes a linear evolution of PMD along the length of the fiber.

We solve the Manakov-PMD equation numerically using the SSFM by splitting the fiber into sections of length $L_\c$. At each step, following the SSFM assumption, we apply the linear and non-linear distortions independently \cite{ssfm}. For a step of length $L_\c$, the CD effect is applied as\footnote{For brevity, we only show the expressions for the X polarization. The expressions for the Y polarization are analogous.}
\begin{equation}
    \Tilde{E}_\x(z+L_\c,f) = \Tilde{E}_\x(z,f) \exp{(\j 2 \beta_2 \pi^2 f^2 L_\c)},
\end{equation}
where $\Tilde{E}_\x(z,f)$ 
is the Fourier transform of $E_\x(z,t)$, 
and $f$ is the baseband frequency. The signal attenuation for each step is applied as
\begin{equation}
    E_\x(z+L_\c,t) = E_\x(z,t) \exp{(- \alpha L_\c / 2)}.
\end{equation}
The Kerr nonlinearity is dependent upon the energy of the baseband signal of both polarizations \cite{kerr}. For dual-polarized transmission, it can be applied in the time domain as
\begin{equation}
\begin{split}
     &E_\x(z+L_\c,t) = \\&\hspace{8mm} E_\x(z,t) \exp{\left(-\j \frac{8}{9} \gamma (|E_\x(z,t)|^2 + |E_\y(z,t)|^2) L_{\eff}\right)},
\end{split}
\end{equation}
where
\begin{equation}
    L_{\eff} = \int_0^{L_\c} \exp{(-\alpha z)} dz = \frac{1 - \exp{(-\alpha L_\c)}}{\alpha}
\end{equation}
is the effective nonlinear step length accounting for the attenuation of signal along the length of the step. Following the simplified polarization impairment model \cite{Zheng:18}, the PMD combined with the rotation of SOP (RSOP) for the $k^{\mathrm{th}}$ step is applied as the rotation of the PSP followed by the DGD operator as
\begin{equation}
    \Tilde{\ve{E}}(z+L_\c,f) =  \Tilde{\ve{E}}(z,f) \ve{R}(\bm{\alpha}^{(k)}) \ve{D}^{(k)}(f),
\end{equation}
where
\begin{equation} \label{DGD}
    \ve{D}^{(k)}(f) = \begin{bmatrix} \exp{(\j\pi f \tau^{(k)} )} & 0 \\ 0 & \exp{(-\j\pi f \tau^{(k)} )}  \end{bmatrix}
\end{equation}
for DGD $\tau^{(k)}$ at the $k^{\mathrm{th}}$ step, and $ \ve{R}(\bm{\alpha}^{(k)}) \in SU(2)$ is the PSP rotation matrix. The rotation matrix can be formulated as
\begin{equation} \label{rotation_matrix}
    \ve{R}(\bm{\alpha}^{(k)}) = \exp{(-\j \bm{\alpha}^{(k)} \cdot \overrightarrow{\bm{\sigma}})  )},
\end{equation}
where $\bm{\alpha}^{(k)} \in \mathbb{R}^3$ and $\overrightarrow{\bm{\sigma}} = (\bm{\sigma}_1, \bm{\sigma}_2, \bm{\sigma}_3)$ is a tensor of the three Pauli spin matrices
\begin{equation}
    \bm{\sigma}_1 = \begin{bmatrix} 1 & 0 \\ 0 & -1  \end{bmatrix}, \bm{\sigma}_2 = \begin{bmatrix} 0 & 1 \\ 1 & 0  \end{bmatrix}, \bm{\sigma}_3 = \begin{bmatrix} 0 & -j \\ j & 0  \end{bmatrix}.
\end{equation}
Note that there can be other formulations of the PSP rotation matrix which belong to the special unitary group of degree~2. Rotation matrices for each step are generated randomly such that they are uniformly distributed over the surface of the Poincar\'e sphere. Different PMD realizations can be generated by choosing a different set of randomly generated DGD and PSP rotation matrices \cite{hinge-model}.

\subsection{Hinge Model for Continuous Polarization Drift} \label{hinge}
In practical fibers, the PMD may drift over time due to environmental factors such as temperature, pressure, cable orientation, stress, cable bends, vibrations, etc \cite{poole1997polarization}. As opposed to the classical PMD literature which assumes all sections to drift, field measurements from various experiments suggest that temporal changes arise from a relatively small number of ``points of stress" in the fiber which may be exposed to the environment \cite{hinge-model2}. This gives rise to the hinge model, where most of the SOP rotation matrices are considered to be static while only the few polarization scramblers at the ``points of stress", referred to as hinges, drift over time. Each hinge $\ve{J}(t) \in SU(2)$ can be formulated as a waveplate and the evolution of $\ve{J}(t)$ over time can be modelled as \cite{hinge-model}
\begin{equation}\label{e:hinge}
    \ve{J}(t) =   \ve{J}(\bm{\alpha}(t)) \ve{J}(t-1),
\end{equation}
where $\ve{J}(\bm{\alpha}(t))$ is a randomly generated innovation matrix of the same form as the PSP rotation matrix defined in equation \eqref{rotation_matrix}. The parameters of the innovation matrix are independently drawn from the following Gaussian distribution for each time instance \cite{hinge-model}:
\begin{equation}
    \bm{\alpha}(t) \sim \mathcal{N}(0, \sigma^2_\p \ve{I}_3),
\end{equation}
where $\sigma^2_\p = 2 \pi \Delta p T$, $\Delta p$ is referred to as the polarization linewidth and $T$ is the time duration between updates. To incorporate continuous PMD drift in the split-step method, we introduce a small number of time-varying hinge matrices at regular intervals across the length of the fiber. In our simulation, we add one hinge at the end of each span considering that the environmental effects would be more prominent near the optical amplifier.

\begin{figure*}[!htbp]
     \begin{subfigure}{0.64\textwidth}
         \centering
         \includegraphics[width=\textwidth]{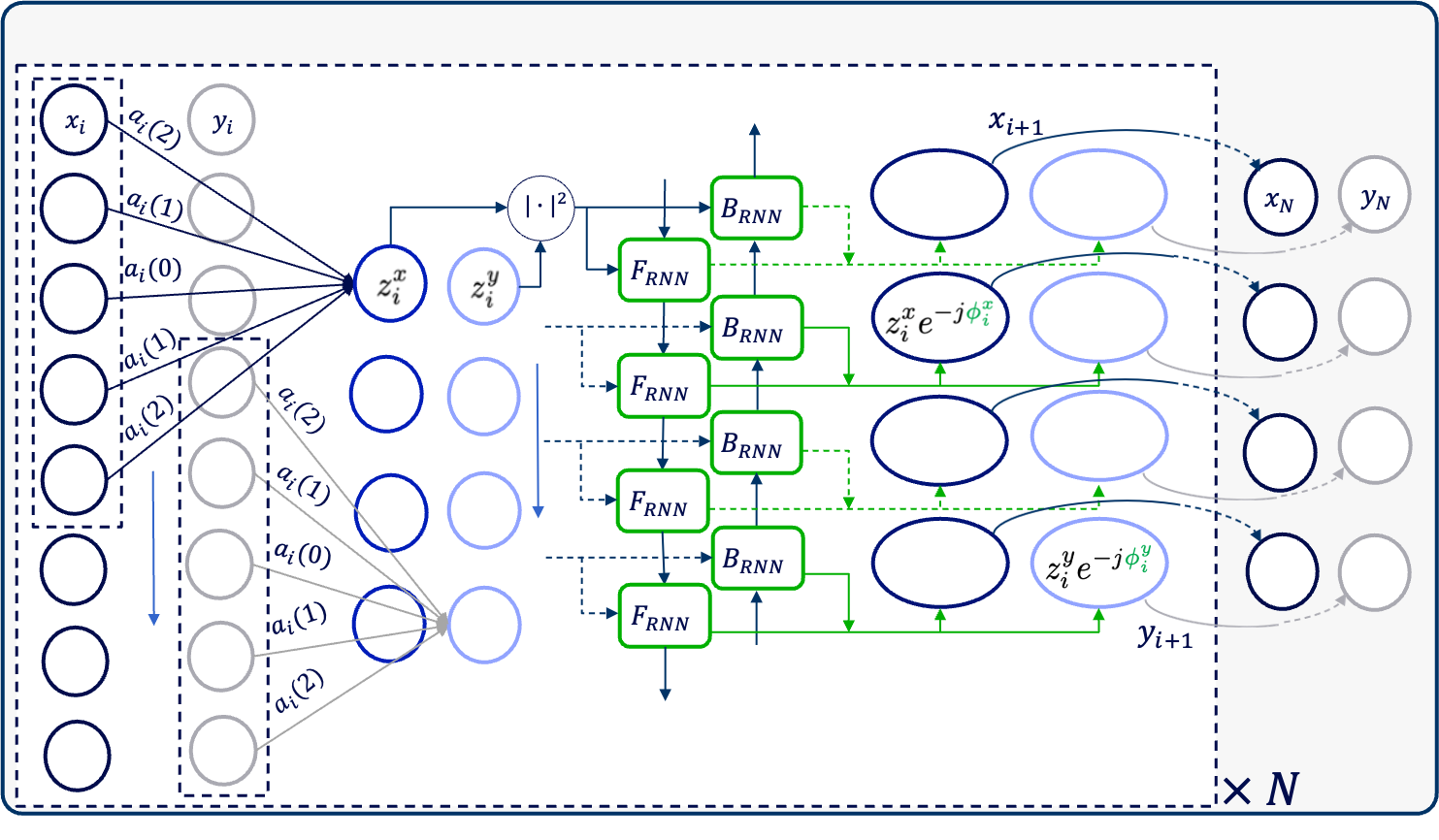}
         \phantomsubcaption
         \label{fig:DCRNN}
     \end{subfigure}
    \hfill
    \begin{subfigure}{0.35\textwidth}
         \centering
         \includegraphics[width=\textwidth]{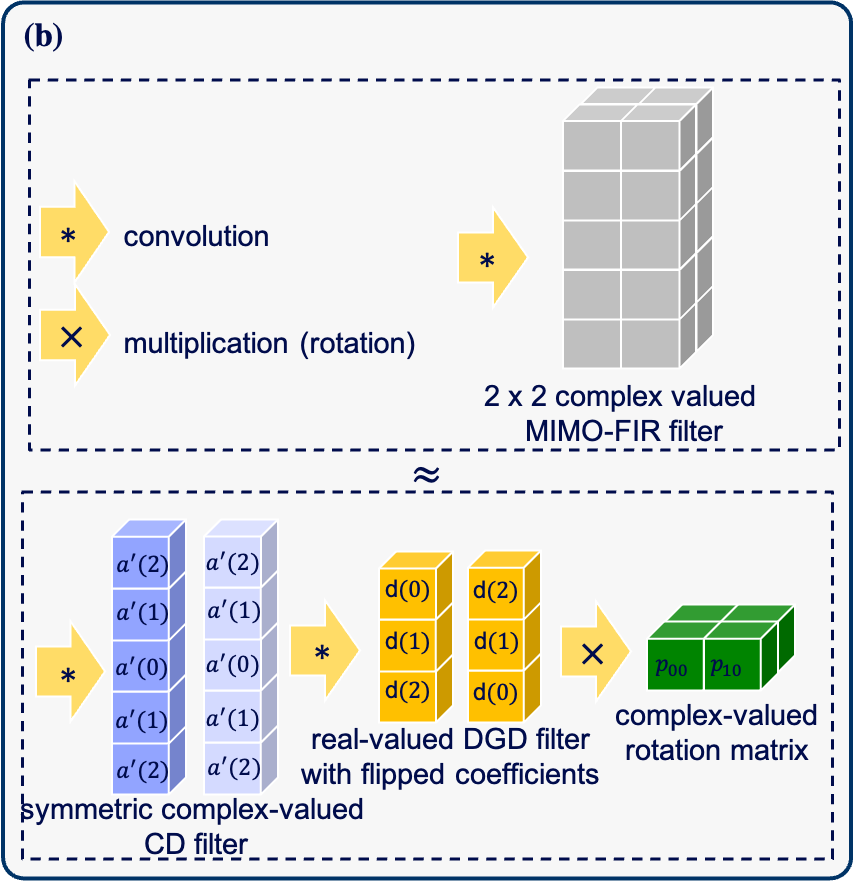}
         \phantomsubcaption
         \label{fig:Dist-PMD}
     \end{subfigure}
     \vspace{-1mm}
    \caption{(a) Architecture of the proposed DCRNN model with $N$ steps. The lumped PMD compensation filter (not shown) is applied to the neural network outputs $(x_N(k), y_N(k))$. (b) For DCRNN-PMD, the linear step of DCRNN is amended to include distributed PMD compensation, and the overall MIMO-FIR filter is decomposed into three steps. }
\end{figure*}

\section{Machine Learning For Joint Compensation of Nonlinearity and PMD} 
\label{s:DCRNN}
In this section, we derive the DCRNN-PMD model with distributed compensation of PMD and present the proposed selective adaptation scheme to efficiently transfer knowledge from an offline trained model, including the incorporation of online learning to develop a real-time adaptation technique. 

\subsection{Deep Convolutional Recurrent Neural Network}
For clarity, in the following we use DCRNN and DCRNN-PMD to differentiate the  variants of the proposed model with lumped and distributed PMD compensation, respectively. Additionally, we use DCRNN(-PMD) when referring to common attributes of both models. 
Fig.~\ref{fig:DCRNN} shows the structure of the DCRNN model for compensating the various impairments. In addition, the DCRNN-PMD model also includes the decomposed form of the modified linear filter depicted in Fig.~\ref{fig:Dist-PMD}.

\subsubsection{Learned Chromatic Dispersion Compensation}
In the DCRNN(-PMD) models, we follow the time domain implementation of CD compensation (CDC) using a finite impulse response (FIR) filter, since it is more efficient \cite{LDBP}. We implement the FIR filter using a one-dimensional (1D) complex-valued convolutional layer. Based on our understanding of the CD effect, as discussed in the previous section, we use the same weights within the convolutional layer for both polarizations. Also, to save computational complexity, we make the convolutional filters symmetrical, which is consistent with the physical model. Some previous works have considered initialization of the convolutional layer based on the associated analytical solution for either constant-linear or mod-logarithmic step size \cite{ldbp-pmd, DCNN}. Although such initialization results in faster convergence, we suspect that it may have a tendency of getting trapped at a local optima during the training process.  Therefore, we initialize the filter weights randomly by sampling values from a uniform distribution $\mathcal{U}(-\sqrt{2l+1},\sqrt{2l+1})$, where $(2l + 1)$ is the width of the convolutional kernel.
The output of the $i^{\mathrm{th}}$ convolution step can be written as
\begin{equation} \label{e:CD}
\begin{split}
    \Tilde{z}^\x_i (k) = \sum_{n = -l}^{l} a_i (n) x_i (k+n),\\
    \Tilde{z}^\y_i (k) = \sum_{n = -l}^{l} a_i (n) y_i (k+n),
\end{split}
\end{equation}
where 
$a_i (n)$ are trainable complex-valued weights at the $i^{\mathrm{th}}$ step and $(x_i(n), y_i(n))$ are the output symbols of X and Y polarizations from the $(i-1)^{\mathrm{th}}$ step, respectively, starting with the received symbols at the first step. In the DCRNN model, the output of the CD compensation layer $ \Tilde{z}^{\x/\y}_i (k)$ is directly input to the RNN based NLC layer as $z^{\x/\y}_i (k)$ (see Fig.~\ref{fig:DCRNN}). At the end of the DCRNN, a two-dimensional (2D) complex-valued convolutional output layer is added to mimic a $2\times2$ multiple-input multiple-output (MIMO)-FIR filter for lumped PMD compensation.

\subsubsection{Distributed Compensation of PMD} \label{dist_pmd}
In the DCRNN-PMD model, a second linear compensation is added at each step for distributed PMD compensation. Based on our discussion of the fiber channel model in the previous section, it can be decomposed into two parts, DGD and rotation of PSP. We follow the design of DGD filter proposed in \cite{ldbp-pmd}, using short real-valued FIR filters, implemented as 1D convolutions. Based on the description of DGD in \eqref{DGD}, we use the same weights for both polarizations but in ``flipped'' order. The application of the DGD filter at the $i^{\mathrm{th}}$ step can be written as
\begin{equation} \label{e:DGD}
\begin{split}
    \hat{z}^\x_i (k) = \sum_{n = -m}^{m} d_i (n)  \Tilde{z}^\x_i  (k+n),\\
    \hat{z}^\y_i (k) = \sum_{n = -m}^{m} d_i (-n) \Tilde{z}^\y_i (k+n),
\end{split}
\end{equation}
where $(2m + 1)$ is the width of the DGD filter and $d_i (n)$ are trainable real-valued weights. This is followed a $2 \times 2$ complex-valued PSP rotation matrix applied as
\begin{equation} \label{e:PSP}
    \begin{bmatrix} z^\x_i (k) \\ z^\y_i (k)  \end{bmatrix} = \begin{bmatrix} p_i^{00} & p_i^{01} \\ p_i^{10} & p_i^{11}  \end{bmatrix} \begin{bmatrix} \hat{z}^x_i (k) \\ \hat{z}^y_i (k)  \end{bmatrix}.
\end{equation}

Fig.~\ref{fig:Dist-PMD} shows the decomposition of the $2\times2$ MIMO-FIR filter, which compensates for all linear impairments, into the component layers for CD, DGD and PSP rotation as described in (\ref{e:CD}), (\ref{e:DGD}) and (\ref{e:PSP}) respectively.
We find that the best performance is obtained when we use free weights for the PSP matrices. However, initializing each PSP rotation matrix as identity matrix $\ve{I}_2$ has been shown to slightly improve the convergence speed of the model \cite{ldbp-pmd}. Other linear impairments such as signal attenuation are assumed to be handled implicitly within the learned CD and PMD compensation layers. 

While the use of free weights in the linear compensation layers provides the best performance, it should be highlighted that different random initializations may yield different performance levels from the trained NN model. In this paper, we present and compare the best performance achieved by each model over multiple random initializations.  

\subsubsection{RNN Based Nonlinearity Compensation} \label{s:rnn}
As discussed in the previous section, the nonlinear phase depends on the energy of the signal from both polarizations. However, for large step size, it is insufficient to simply consider the energy of samples at the current time-step. Rather, the energy of all samples within the dispersion spread of that step must be considered to account for the interplay of dispersion and nonlinearity \cite{essfm}. In the DCRNN(-PMD) models, we introduce a novel bidirectional recurrent neural network (BiRNN) layer which efficiently processes the energy of past (and future) samples as a sequence by preserving the information in its memory. The incorporation of the BiRNN layer presents a generalized parameterization of the low-pass filtered DBP \cite{du_lowery:2014}. To save computational complexity, we process the received signal in large blocks and only consider the output of the center RNN cells from each direction. This cost saving strategy is similar to the center-oriented long short term memory (Co-LSTM) model \cite{co-lstm} except that we only use ordinary RNN cells to reduce cost further by a factor of 4. The nonlinear phase corrections $\phi_i^{\x}$ and $\phi_i^{\y}$, highlighted in Fig.~\ref{fig:DCRNN}, are estimated as
\begin{equation}
\begin{split}
    \phi_i^{\x}(k) = (\ve{f}^{\x}_i)^T \ve{h}_{\f,i}(k)  + (\ve{b}^{\x}_i)^T \ve{h}_{\b,i}(k),\\
    \phi_i^{\y}(k) = (\ve{f}^{\y}_i)^T \ve{h}_{\f,i}(k)  + (\ve{b}^{\y}_i)^T \ve{h}_{\b,i}(k),
\end{split}
\end{equation}
where
\begin{equation}
\begin{split}
  \!\!  &\ve{h}_{\f,i}(k) =\tanh(\ve{W}_{\f,i} [\ve{h}^T_{\f,i}(k\!-\!1), |z^{\x}_i (k)|^2,|z^{\y}_i (k)|^2]^T),\\
  \!\!  &\ve{h}_{\b,i}(k) = \tanh(\ve{W}_{\b,i} [\ve{h}^T_{\b,i}(k\!+\!1), |z^{\x}_i (k)|^2,|z^{\y}_i (k)|^2]^T),
\end{split}
\end{equation}
and $\ve{W}_{\ell,i}\in\mathbb{R}^{n_h\times (n_h + 2)}$, $\ell\in\{\f,\b\}$, and $\ve{f}^{p}_i\in\mathbb{R}^{n_h}$, $\ve{b}^{p}_i\in\mathbb{R}^{n_h}$, $p\in\{\x,\y\},$ 
 are trainable real-valued weights, and $n_h$ is the number of hidden units in each RNN. Each step ends with the application of NLC as
\begin{equation}
\label{e:NLCstep}
\begin{split}
    x_{i+1}(k) = z_i^\x(k) \e^{-\j\phi_i^\x(k)},\\
    y_{i+1}(k) = z_i^\y(k) \e^{-\j\phi_i^\y(k)}.
\end{split}
\end{equation}
Our results in Section~\ref{results} suggest that ordinary RNN cells are sufficient to capture the interplay of dispersion and nonlinearity for the length of the step. 

As indicated in Fig.~\ref{fig:DCRNN}, the described compensation steps are repeated $N$ times, i.e., $i=0,1,\ldots, N-1$.

\subsection{Transfer Learning Based Continuous Adaptation} \label{s:adatationscheme}
To perform joint compensation of nonlinearity and PMD, it is essential that the learned model is able to adapt to changes in PMD. The current state of PMD in a fiber link may be different from that of the training data on which the model is initially trained. 

\subsubsection{Retraining}
It is prohibitively expensive to recollect the required training data and rebuild the model. Since there could be infinitely many PMD realizations, even in links drawn from the same spool of fiber \cite{hinge-model2}, it is essential that we minimize the retraining effort for every installation. Therefore, we turn our attention towards transfer learning (TL) \cite{TL}. The important question in TL is ``what to transfer?''. In the DCRNN(-PMD) models introduced above, we can clearly interpret the function of each layer based on the model design. 
In theory, the parameters of the CD and NLC layers should not require adaptation if only the PMD effect is different. Therefore, we can reuse (transfer) the associated learned coefficients without any adaptation. In \cite{ldbp-pmd}, the CD parameters from the pre-trained LDBP model are similarly transferred to initialize the LDBP-PMD model. In our proposed selective training scheme, we freeze the CD and NLC layers and only re-train the parameters associated with PMD compensation. Consequently, for lumped PMD compensation, only the output layer of the model needs to be re-trained. For distributed PMD compensation schemes, i.e., DCRNN-PMD, only the DGD filter and the PSP rotation matrices within each step need to be updated. In Section~\ref{results}, we show that this significant reduction in trainable parameters greatly reduces the re-training efforts for knowledge transfer. 

\subsubsection{Tracking}
In contrast to \cite{ldbp-pmd}, we consider continuous evolution in the state of PMD, where each successive state is related to the previous state as described in (\ref{e:hinge}).
In order to continuously adapt the model in real-time, we combine the transfer learning solution with online learning in the proposed adaptive scheme. We divide the training process into two parts: acquisition and tracking. For acquisition, we send pilot symbols through the channel to fine-tune the model parameters until the performance of the filter stabilizes. Considering the associated overhead, during normal operation, we can only send a very limited number of pilots in each transmitted block which may be insufficient for sustained model adaptation. Therefore, after acquisition, we move into a decision directed tracking stage where we only send information symbols. The learned filter performs equalization in small blocks. At the tracking stage, the hard decoded output symbols from the previous processed block are used to update the filter in operation. Note that the choice of block size is directly related to the delay in feedback to the model. For continuous adaptation, a small feedback delay (FD) is desired to avoid mismatch between the training and test distributions. Consequently, a trade-off emerges between the feedback delay and the computational complexity of the model for which, as discussed in Section \ref{s:rnn}, a larger block size is desirable. This trade-off has not been studied in the present work and would be a crucial line of inquiry for future investigations. Another key difference in the online training routine is the use of stochastic gradient descent (SGD) optimizer without momentum as opposed to the \textit{Adam} optimizer employed predominantly in offline training. We find that the use of momentum negatively impacts the online training process. This is because the momentum term tries to push the gradient in the same direction as the previous iterations, which is no longer warranted since the underlying gradient surface is continuously evolving due to PMD drift. In the next section, we discuss the methods used to analyze the performance-complexity profile of the learned NLC schemes.

\begin{figure*}
    \centering
    \includegraphics[width=0.85\textwidth]{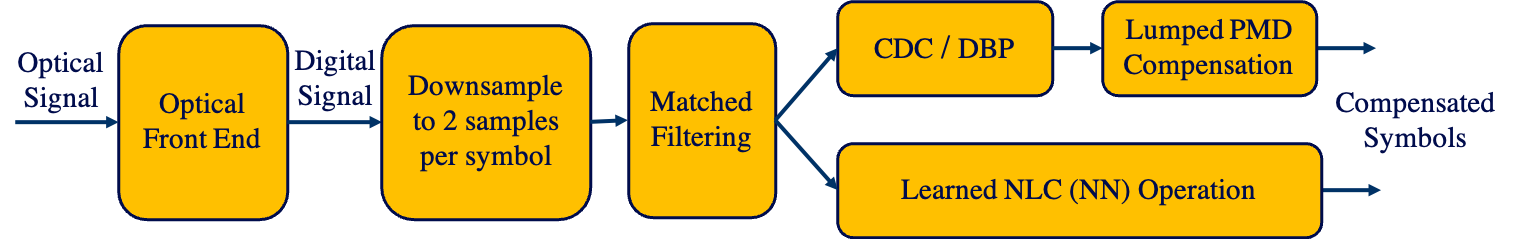}
    \caption{Receiver processing schematic for various NLC schemes.}
    \label{fig:receiver}
\end{figure*}

\section{Performance-Complexity Analysis of\\ Neural Network Based NLC Schemes}
\label{pruning}
In order to compare the computational complexity of learned and deterministic NLC techniques, it is important to acknowledge that various learned models could have different levels of redundancy due to the presence of unimportant parameters within the neural networks.
Therefore, for a fair comparison and practical application, it is essential that we reduce the complexity of each model down to its optimal level. 
For neural networks, model compression can be achieved by weight pruning, which has been shown to be effective for complexity reduction of NN-based NLC \cite{pruning}. We perform pruning as a semi-structured iterative process. Each iteration comprises of two basic steps: pruning and retraining. In the pruning step, we remove $20\%$ of the smallest weights within each layer of the model independently, since it is non-trivial to compare the weights of different layers. For retraining, we apply learning rate rewinding (LRR) based on the lottery ticket hypothesis (LTH) \cite{LTH}. In LRR, we reset the learning rate schedule to an earlier iteration in the training process prior to fine-tuning the pruned network. This mimics the effect of rewinding the pruned model to an earlier training iteration which, as per LTH, allows the model to achieve a better optima.

In the literature, the performance complexity trade-off of various learned NLC models is predominantly examined by training differently sized variants of the model and measuring their performance \cite{lumped_comparison, DCNN, LDBP}. However, in our study, we observed that model performance falls much more sharply if we train a smaller neural network from the beginning. By comparison, it is better to start with an overbuilt neural network and iteratively prune the weights to reduce its complexity. In \cite{Chang2021ProvableBO}, Chang \textit{et al.} identify a double descent behavior, where the risk of the pruned model is consistently improved with overparameterization. Therefore, in this study, we examine the performance complexity trade-off of learned NLC models by measuring the performance of the fully parameterized base model at various iterations of the pruning process. The results obtained from this approach are discussed in Section \ref{s:numericalresults}.

In the case of online learning for tracking continuous changes in PMD, it is also important to consider the complexity of the optimization step performed using backpropagation \cite{Loshchilov2016SGDRSG}. During online training, each sample is only used in a single update iteration, thus, the computational complexity of the backpropagation remains of the same order as that of the forward propagation. Therefore, reducing the number of parameters through pruning directly improves the associated training overhead. The number of computations required are further reduced since the updated weights for non-PMD related layers need not be calculated. During this study, we have assumed that the computation of updated weights during online training can be performed as a parallel process.

\section{Numerical Results and Discussion} \label{results}
In this section, we present simulation results  (i) to highlight the benefits of the proposed DCRNN(-PMD) models in terms of performance-complexity trade-off and (ii) to demonstrate the effectiveness of the proposed transfer and online learning for adjusting to PMD changes. 

\subsection{Simulation Setup}
We test the proposed and benchmark NLC methods on two configurations of a 32~GBd dual-polarized 64-QAM transmission system over  a $12 \times 80$~km standard single-model fiber (SSMF), with a single channel and with 5~WDM channels, consistent with previous works on NN based NLC \cite{CVNN, jain:2022} to facilitate comparison. The detailed  parameters are provided in Table~\ref{tab:fiber_parameters}. We simulate the forward transmission using the SSFM, as described in Section~\ref{SSFM} by dividing the fiber into 500 uniform steps per span (StPS) with an erbium-doped fiber amplifier (EDFA) at the end of each span.  
As discussed in Section~\ref{SSFM}, different choices of DGD and PSP rotation matrices generate different realizations of PMD. We generate data from 8 different PMD realizations to train and test the learned models. At the receiver, the signal is coherently detected and sampled at 2 samples per symbol. A root raised cosine (RRC) matched filter is deployed when performing CDC and its position in the transmission chain is maintained while transitioning to various NLC schemes.

\begin{table}
        \renewcommand{\arraystretch}{1.5}
        \centering
        \caption{Fiber Parameters}
        \begin{tabular}{|c|c|}
            \hline
            Parameter & Value \\\hline
            No. of channels & 1, 5 \\\hline
            Modulation & 64 QAM \\\hline
            Transmission length & 12$\times$80 km \\\hline
            Baud Rate & 32 GBd \\\hline
            Channel spacing & 37.5 GHz \\\hline
            $\alpha$ & 0.21 dB/km \\\hline
            $\beta_2$ & $-$21.49 ps$^2$/km \\\hline
            $\gamma$ & 1.14 /W-km\\\hline
            Noise figure & 4.5 dB \\ \hline
            PMD & 0.1 ps$/\sqrt{\mathrm{km}}$ \\ \hline
            SSFM slices per span & 500 \\\hline
        \end{tabular}
        \label{tab:fiber_parameters}
\end{table}

Fig.~\ref{fig:receiver} depicts the processing at the receiver for various learned and deterministic NLC methods considered for comparison. Downsampling to symbol space is performed implicitly within the output layer of the learned NLC solution. For deterministic methods, it is performed by the lumped PMD compensation filter described in the following section. Next, we describe the implementation and (hyper)parameters of the various NLC schemes applied.

\subsection{Setup of NLC Methods}
\label{s:training}

\subsubsection{Baseline Deterministic Methods}
We compare the learned model against conventional DBP \cite{DBP} with uniform step size at 1, 2 and 3 StPS using 2 samples per symbol. At each step, the CDC is performed using a frequency domain equalizer followed by the NLC step applied as
\begin{equation}
\begin{split}
     \hat{E}_\x(t) = \Tilde{E}_x(t) \exp{(-\j \xi \gamma \Lambda (|\Tilde{E}_\x(t)|^2 + |\Tilde{E}_\y(t)|^2) L_{\eff})},\\
     \hat{E}_\y(t) = \Tilde{E}_\y(t) \exp{(-\j \xi \gamma \Lambda (|\Tilde{E}_\x(t)|^2 + |\Tilde{E}_\y(t)|^2) L_{\eff})},
\end{split}
\end{equation}
where $(\Tilde{E}_\x(t), \Tilde{E}_\y(t))$ are the output of the CDC step,  $\Lambda$ is the relative power scaling factor to account for signal attenuation (amplification in case of backward propagation) over the step, and $\xi$ is a scaling coefficient numerically optimized for each launch power. To compensate for PMD, a least mean squares (LMS) based adaptive $2 \times 2$ MIMO-FIR filter is applied after the DBP operation with 19 taps, in accordance with the PMD parameter.

\subsubsection{Baseline Learned Methods}
Among the learned NLC models, we consider LDBP with both lumped and distributed compensation of PMD \cite{LDBP, ldbp-pmd}. In both cases, we consider a model with 1~StPS operating at 2~samples per symbol. All the complex-valued convolutional CD filters have the same length with 29 taps each, chosen based on the dispersion spread of the fiber. For lumped PMD compensation, we add an additional 2D complex-valued convolutional layer at the end, identical to the $2 \times 2$ MIMO-FIR filter used for deterministic methods. The complete model, including the lumped PMD compensation layer is trained together. For distributed PMD compensation, we include DGD filters and rotation matrices at each step, as described in Section~\ref{dist_pmd}. All real valued DGD filters have the same length with 3 taps each and the same weights are used for both polarizations but in inverted order.

\subsubsection{Proposed DCRNN Model} For the proposed DCRNN(-PMD) models, we again consider both lumped and distributed PMD compensation to highlight the associated performance gain. For a consistent comparison, we use the same kernel width for the complex valued convolutional CD filter and the real valued DGD filter as that used in the LDBP-PMD model. For the BiRNN based NLC layer, we use just two ordinary RNN units each in both the forward and backward RNN. 

\subsubsection{Training of Learned Models}

All learned models are implemented using PyTorch \cite{pytorch} and initially trained offline using the \emph{Adam} optimizer \cite{Adam} of stochastic gradient descent with a cosine annealed learning rate schedule \cite{Loshchilov2016SGDRSG}. The learning rate for each iteration can be defined as
\begin{equation}
    \eta_n = \eta_{\min} + \frac{1}{2}(\eta_{\max}-\eta_{\min}) \left( 1 + \cos{ \left( \frac{n}{N_\mathrm{iter}}\pi  \right) }\right),
\end{equation}
where $\eta_n$ is the learning rate for the $n^{\mathrm{th}}$ training iteration, $N_{\mathrm{iter}}$ is the total number of iterations, and $\eta_{\min}$ and $\eta_{\max}$ are the minimum and maximum learning rates respectively. We switch to the SGD optimizer without momentum as per the proposed online training scheme discussed in Section \ref{s:adatationscheme} with a constant learning rate to adapt the offline trained model under PMD drift. The complete list of training parameters are reported in Table~\ref{tab:training_parameters}.

\begin{table}[t]
    \renewcommand{\arraystretch}{1.5}
    \centering
    \caption{Training Parameters}
    \begin{tabular}{|c|c|}
        \hline
        Parameter & Value \\\hline
        Training symbols & $2^{19}$ \\\hline
        Optimizer & Adam \\\hline
        Learning rate  schedule & Cosine annealing \\\hline
        Learning rate parameters & $\eta_{\max}=10^{-3}$, $\eta_{\min}=10^{-5}$\\\hline
        Test symbols & $2^{19}$ \\\hline
        Epochs & $10^3$ \\\hline
        Batch size & $10^4$ \\\hline
        Validation split &  $80:20$ \\\hline
    \end{tabular}
    \label{tab:training_parameters}
\end{table}

\subsection{Numerical Results} \label{s:numericalresults}
We measure the performance of each method based on the Q-factor \cite{qfactor} for the channel of interest (CoI) calculated from the BER as
\begin{equation}
   Q[\mathrm{dB}] = 20 \log_{10}{\left( \sqrt{2}\ \mathrm{erfc}^{-1}(2\mathrm{BER}) \right)}.
\end{equation}
The BER is obtained by directly counting the errors. We evaluate the performance gains of each NLC scheme against that of linear compensation only using a lumped frequency domain CD equalizer followed by a lumped LMS based MIMO-FIR PMD filter.

\begin{figure}[t]
     \centering
     \includegraphics[width=\linewidth]{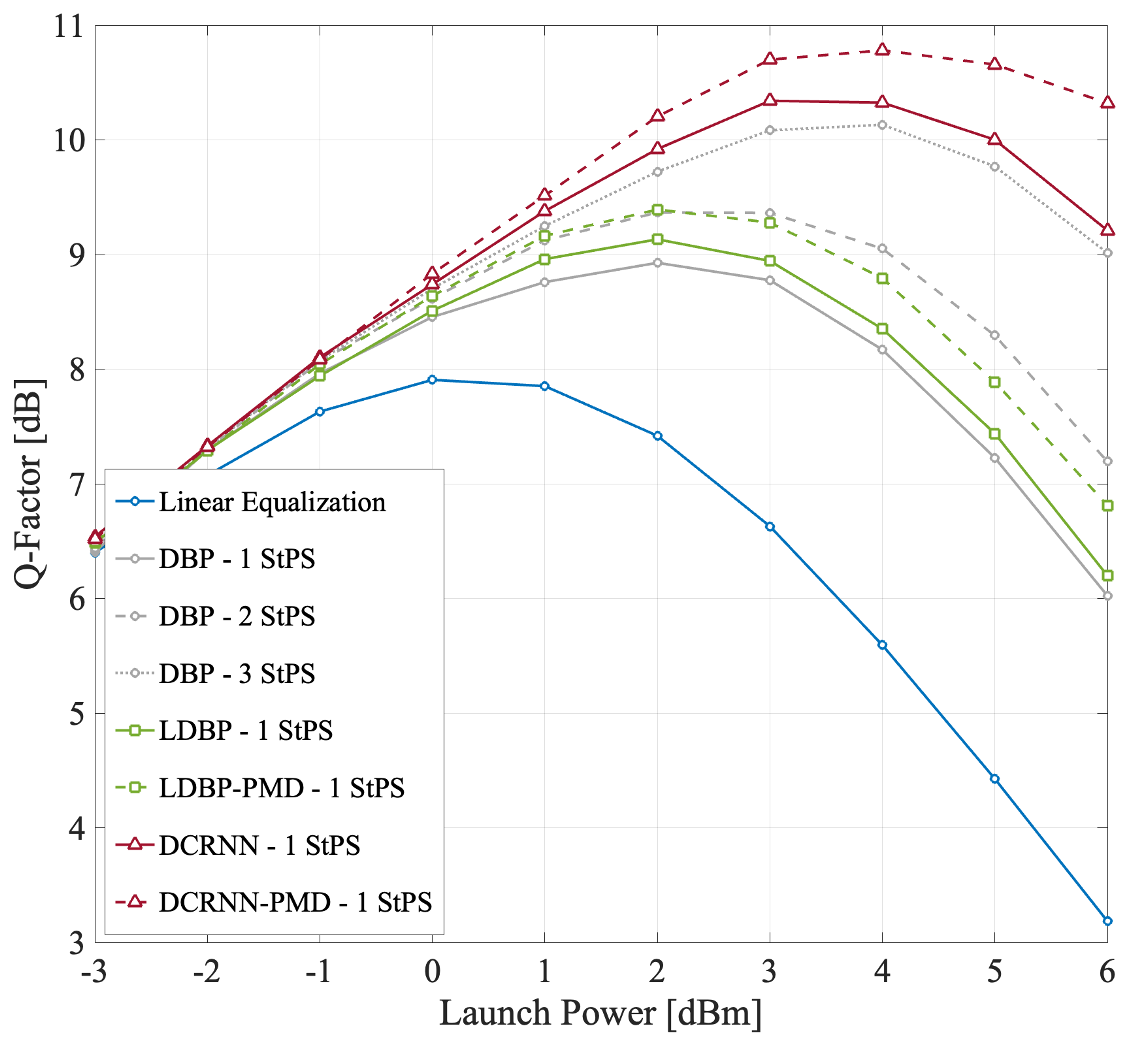}
     \caption{Performance of NLC schemes for single channel.
     }
     \label{fig:DistributedSC}
\end{figure} 
\begin{figure}[t]
     \centering
     \includegraphics[width=\linewidth]{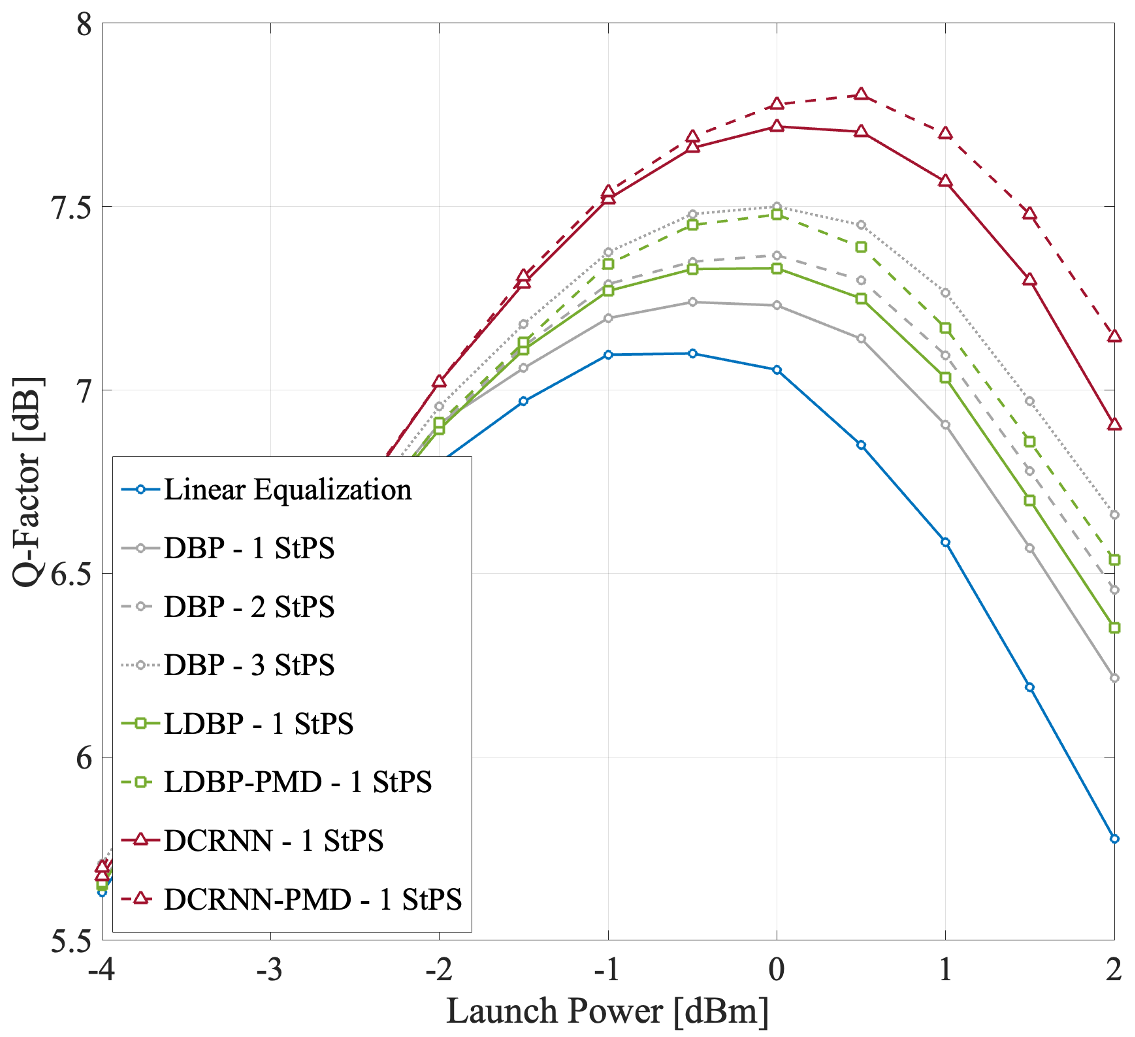}
     \caption{Performance of NLC schemes for 5-WDM channel.
     }
     \label{fig:DistributedWDM}
\end{figure}
\subsubsection{Static Channel}
We first look at the performance of each method in case of a static channel without considering time evolution of PMD. Fig.~\ref{fig:DistributedSC} and Fig.~\ref{fig:DistributedWDM} show the Q-factor as a function of launch power for single channel and WDM transmission, respectively. For the single channel case, the proposed DCRNN model with only lumped PMD compensation delivers state of the art performance by  outperforming all baseline learned and deterministic NLC schemes. It provides a $2.43$~dB gain over linear compensation, $1.42$~dB gain over DBP at 1-StPS, and $1.21$~dB gain over LDBP at 1-StPS. It even outperforms LDBP-PMD with distributed compensation of PMD by $0.99$~dB. The results show a strong impact of the interaction between nonlinearity and dispersion. Remarkably, using just 2 ordinary RNN cells in each BiRNN layer, the DCRNN model is able to capture and compensate for this interaction.\footnote{We note that it may be prudent to re-examine the performance complexity trade-off of using different kinds of RNN cells for the case that the dispersion spread for the step widens due to, e.g., increasing baud rates or simply larger steps.} Furthermore, using distributed PMD compensation, the DCRNN-PMD model provides another $0.44$~dB Q-factor gain. The performance of the LDBP model also improves by $0.26$~dB after including distributed PMD compensation, depicted as LDBP-PMD. This demonstrates the benefit of joint distributed NLC and PMD compensation.  Increasing the number of taps in the DGD filter does not result in any noticeable improvement in performance, which is expected considering the small amount of DGD per step. 

The performance gains for all schemes in the WDM case are markedly slimmer than the single channel case due to the presence of cross phase modulation (XPM) \cite{Dar:2017}. Note that none of the models considered in this study explicitly mitigate the XPM effects. However, we still observe a similar trend in the relative performance of various NLC schemes. For the CoI, the DCRNN-PMD model again outperforms all other learned and deterministic methods achieving $0.71$~dB Q-factor gain over linear compensation, $0.55$~dB gain over DBP at 1-StPS, and $0.38$~dB gain over LDBP at 1-StPS.

\subsubsection{Impact of Iterative Weight Pruning}
\begin{figure}
     \centering
     \includegraphics[width=\linewidth]{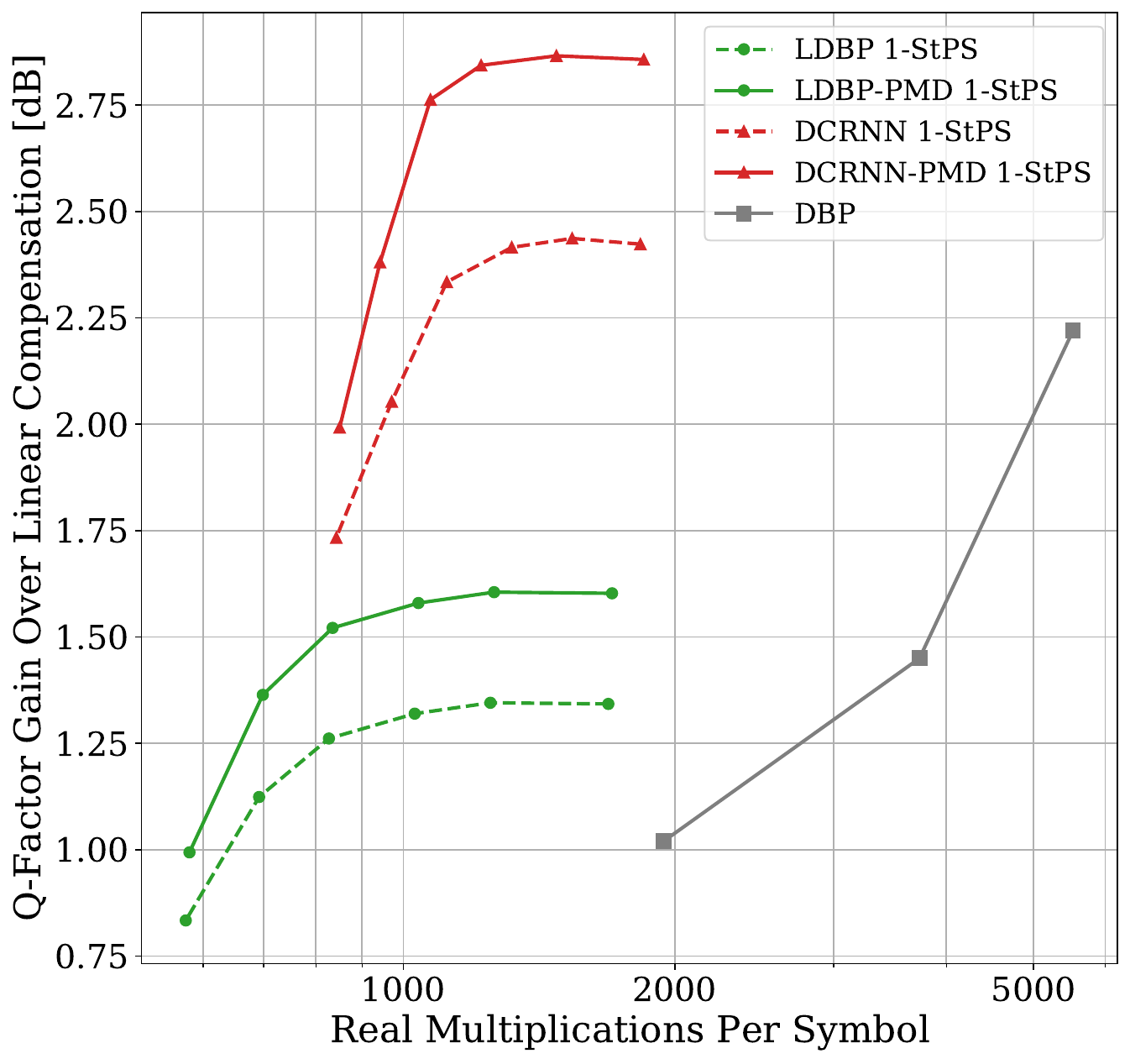}
     \caption{Performance-complexity comparison of  NLC schemes.
     }
     \label{fig:PCComparison}
\end{figure}

In Fig.~\ref{fig:PCComparison}, we show the performance complexity comparison for the learned NLC methods along with conventional DBP for the single channel case. The rightmost point on each curve represents the performance at the hyperparameters provided earlier. Each subsequent point to the left is obtained using the iterative pruning and fine-tuning method discussed in Section~\ref{pruning}. For conventional DBP, different points are generated by using different number of steps per span. We use the number of real multiplications per compensated symbol as a measure of computational complexity. Using the iterative weight pruning, the computational complexity of the considered learned NLC methods can be reduced by up to  $50\%$ without incurring significant performance loss. At 1-StPS, the proposed DCRNN-PMD model continues to provide $1.3$~dB Q-factor gain over DBP even after being pruned to half the computational cost.

\subsubsection{Instantaneous Changes in PMD} \label{s:InstPMD}
In the next set of results, we take the first step towards adapting to changes in PMD. Until this point, the learned models were trained and tested on data collected from the same realization of PMD, i.e., the channel under test was the same as the channel under training. While we continue to assume the channel to be stationary, we now consider that the current PMD realization of the channel may be different from when the data was collected for initial offline training. For this, we collect data from 8 different PMD realizations and train the model with only the data from the first realization using the offline training routine discussed in the Section~\ref{s:training}. Our goal is to transfer knowledge from this trained base model such that the model can be used for other PMD realizations. 
As per the proposed transfer learning based adaptive scheme discussed in Section \ref{s:adatationscheme}, we now restrict the re-training to only the PMD compensation parameters. 

\begin{figure}[t]
    \centering
    \includegraphics[width=\linewidth]{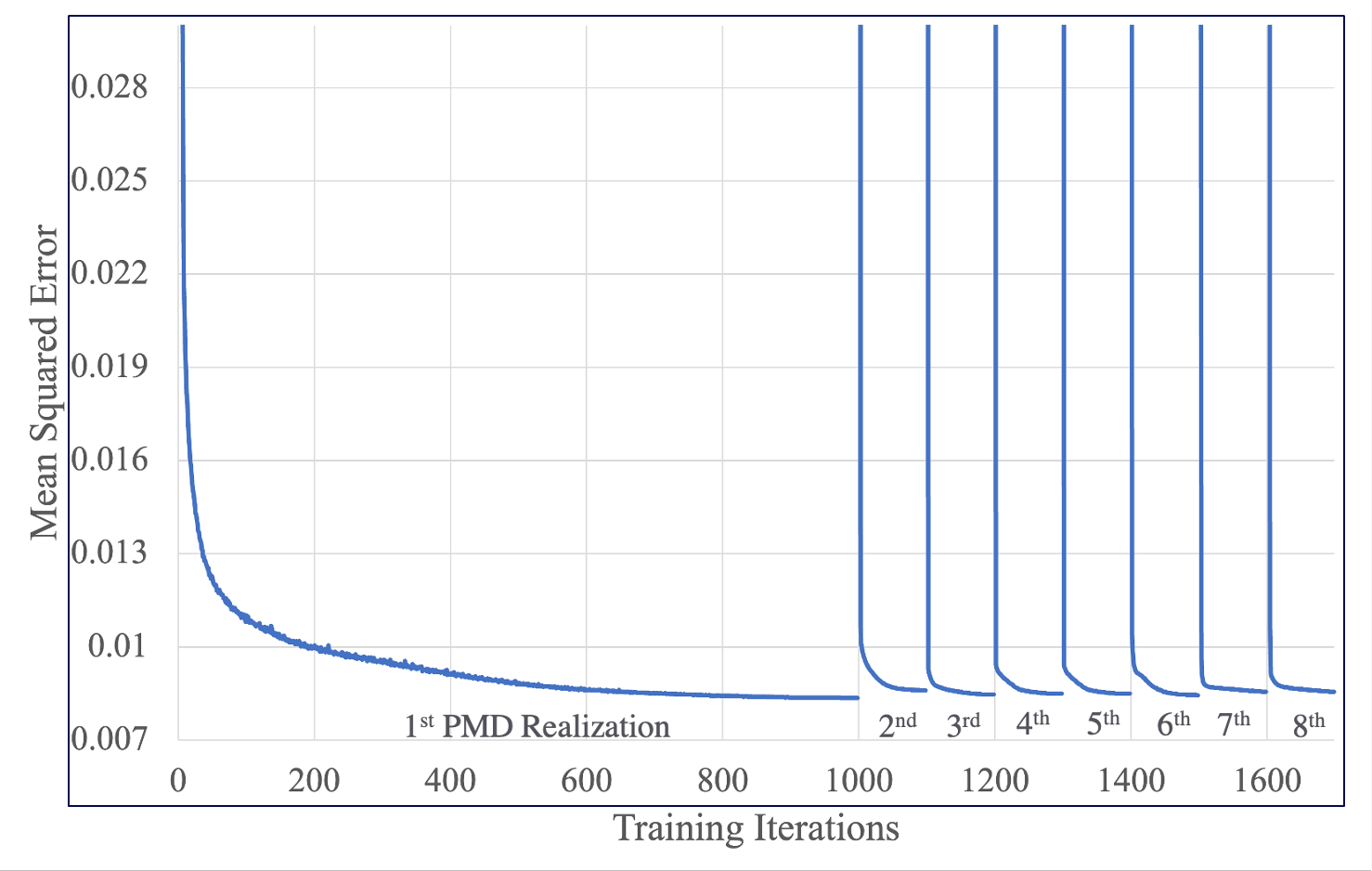}
    \caption{Validation loss for re-training base model for different PMD realizations. The model is  DCRNN-PMD. 
}
    \label{fig:8pmd_mse}
\end{figure}

Fig.~\ref{fig:8pmd_mse} shows the convergence of mean squared error (MSE) for initial training followed by re-training for the 7 remaining PMD realizations for the DCRNN-PMD model. It can be observed that while initial training is performed for $10^3$ epochs, the MSE converges to within $1\%$ of the minimum value in only $50$ re-training epochs on average, representing a $95\%$ reduction in required training time. More importantly, this result proves our conjecture for the proposed transfer learning solution that the model can be adapted by only re-training the parameters associated with PMD compensation.

\subsubsection{Continuous Changes in PMD}
Finally, we discard the stationary assumption for the channel and introduce continuous drift in PMD based on the hinge model from Section~\ref{hinge}. In our simulation, we add one hinge at the end of each span. We believe that this is a good choice since the environmental effects causing PMD rotation are likely to be more prominent near the optical amplifier. 
\begin{figure}[t]
    \centering
    \includegraphics[width=\linewidth]{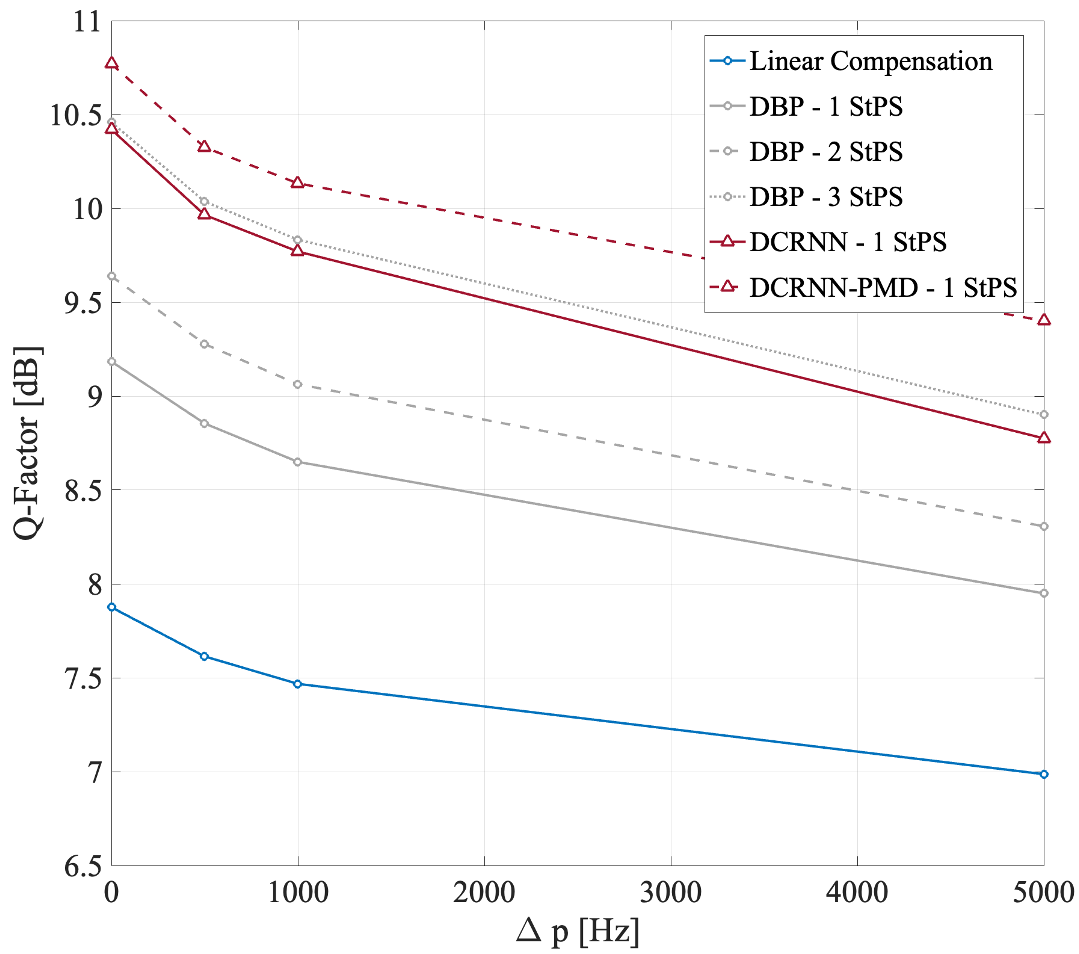}
    \caption{Peak performance of each NLC method for different amount of PMD drift. } 
    \label{fig:SOP_Tracking}
\end{figure}
To set a baseline, we apply linear equalization and DBP with a decision directed LMS based MIMO-FIR filter. The lumped and distributed PMD compensation methods examined have only $152$ and $132$ trainable parameters respectively. We apply the proposed online training routine from Section \ref{s:adatationscheme} with FD $ = 1$.  We measure the tracking performance of each method at four different rates of PMD drift by choosing values of the polarization linewidth $\Delta p$ as $0$~Hz, $500$~Hz, $1000$~Hz and $5000$~Hz, respectively. The learning rate for adapting each NLC method at each value of $\Delta p$ is optimized empirically. In Fig.~\ref{fig:SOP_Tracking}, we show the peak Q-factor achieved by each method at different values of $\Delta p$. The Q-factor is measured at each point in time using a sliding window of $50,000$ symbols. The sliding window needs to be large enough to capture a sufficient number of errors to reliably estimate the Q-factor. The large window averaging also makes the performance curves appear much smoother. 
We observe that performance  reduces with increasing polarization linewidth, as it is expected. However, the proposed DCRNN and DCRNN-PMD models are able to maintain their superiority over conventional DBP and linear equalization. 

\begin{figure}[t]
    \centering
    \includegraphics[width=\linewidth]{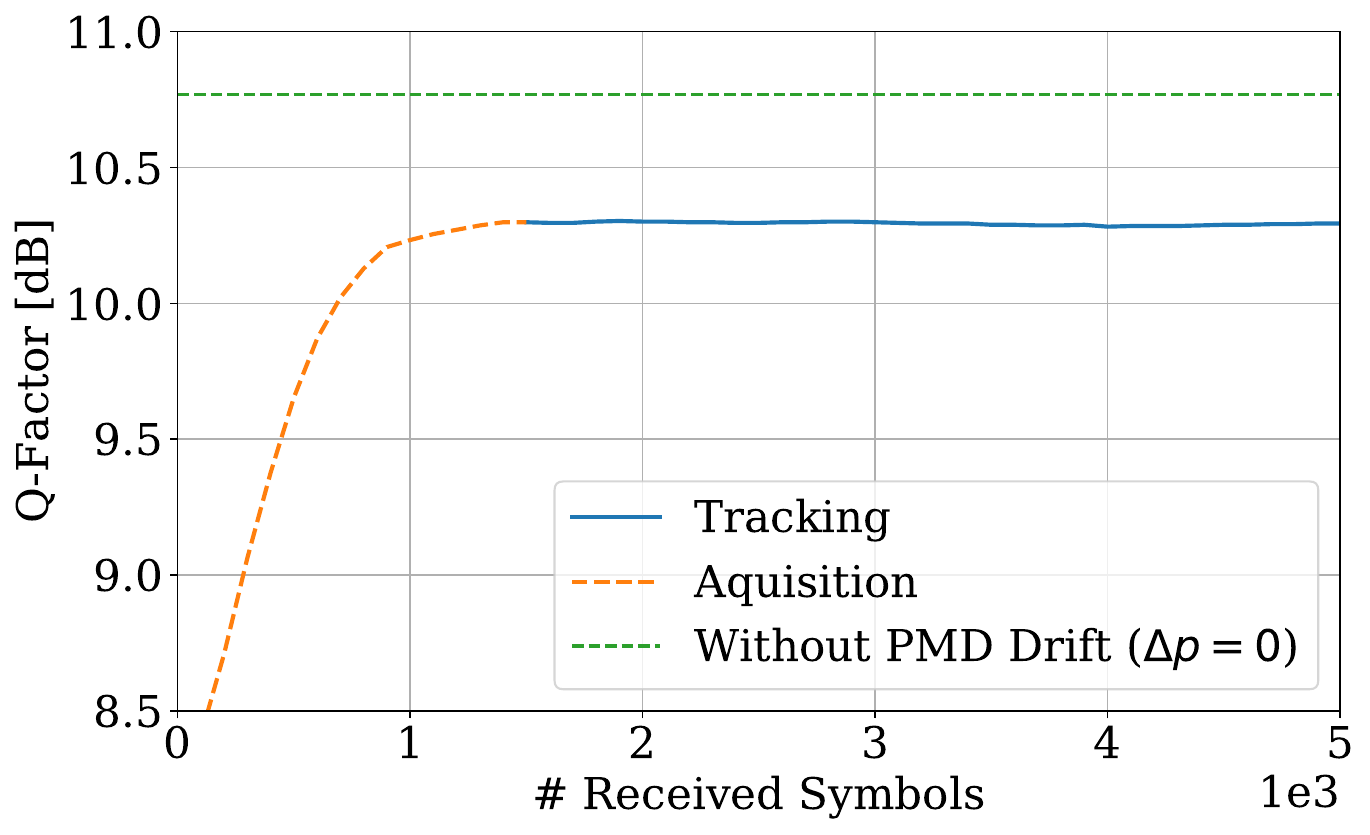}
    \caption{Q-factor trend of DCRNN-PMD model at $\Delta p = 500$ Hz.}
    \label{fig:RSOP_Tracking}
\end{figure}
Fig.~\ref{fig:RSOP_Tracking} provides further insight into the interplay of adjusting to a different PMD realization, i.e., acquisition, and tracking the PMD drift due to SOP changes. We show the running performance of DCRNN-PMD model at $\Delta p = 500$~Hz. Again, the Q-factor at each point in time, represented by the number of received symbols, is calculated using a sliding window. We use this scenario to highlight the remarkable rate at which the DCRNN-PMD model converges to the peak Q-factor using online learning. 

We close our discussion by noting that in scenarios where the error rate (before error correction decoding) is higher, tracking using decision-directed feedback has the potential for error propagation from decision errors. Hence, tracking may experience a performance degradation. This could be mitigated through the use of periodic pilot signals and restricting the decision-directed feedback to only the most reliable decisions. The exact interplay of the use of decision feedback, decision errors, and tracking performance is a subject for future studies.

\section{Conclusion} \label{conclusion}
In this paper, we have discussed the DCRNN-PMD model for NLC with distributed compensation of PMD. We have shown that the DCRNN-PMD model provides notable performance gains over previously proposed learned NLC methods as well as DBP. For example, using iterative pruning,  the proposed model delivers $1.3$~dB Q-factor gain over conventional DBP at only half the computational cost for the considered dual-polarized 64-QAM transmission at 32 GBd over a $12 \times 80$~km SSMF. We have extended the DCRNN-PMD model to facilitate adaptation to PMD changes. The proposed transfer learning based selective training scheme successfully adapts the offline trained model to the current state of PMD in only a fraction of the initial training time. The training effort has further been reduced by extending the  scheme with principles of online learning. Results for a PMD drift use case highlight that combined transfer and online learning enables near real-time tracking of PMD. We thus have demonstrated that our proposed method can be used to integrate a learned NLC solution in practical optical fiber communication systems.

\section*{Acknowledgments}
This research was supported by the Natural Sciences and Engineering Research Council of Canada (NSERC) and Huawei Tech., Canada. It was supported in part through computational resources and services provided by Advanced Research Computing at the University of British Columbia.

\bibliographystyle{IEEEtran}
\bibliography{references}

\end{document}